# Tetrameric UvrD helicase is located at the *E. coli* replisome due to frequent replication blocks

Adam J. M Wollman[1,2,3], Aisha H. Syeda[1,2], Andrew Leech[4], Colin Guy[5,6], Peter McGlynn[2,6], Michelle Hawkins[2] and Mark C. Leake[1,2]

The authors wish it to be known that, in their opinion, the first two authors should be regarded as joint First Authors

[1] Department of Physics, University of York, York YO10 5DD, United Kingdom.

[2] Department of Biology, University of York, York YO10 5DD, United Kingdom.

[3] Current address: Biosciences Institute, Newcastle University, NE1 7RU, United Kingdom.

[4] Bioscience Technology Facility, Department of Biology, University of York, York YO10 5DD, United Kingdom

[5] Current address: Covance Laboratories Ltd., Otley Road, Harrogate, HG3 1PY, United Kingdom

[6] Previous address: School of Medical Sciences, Institute of Medical Sciences, University of Aberdeen, Foresterhill, Aberdeen AB25 2ZD, United Kingdom

\* To whom correspondence should be addressed. To whom correspondence should be addressed. Tel: +44 (0)1904322697. Email: mark.leake@york.ac.uk

Present Address: Departments of Physics and Biology, University of York, York YO10 5DD, United Kingdom

## ABSTRACT

DNA replication in all organisms must overcome nucleoprotein blocks to complete genome duplication. Accessory replicative helicases in *Escherichia coli*, Rep and UvrD, help replication machinery overcome blocks by removing incoming nucleoprotein complexes or aiding the re-initiation of replication. Mechanistic details of Rep function have emerged from recent live cell studies, however, the activities of UvrD *in vivo* remain unclear. Here, by integrating biochemical analysis and super-resolved single-molecule fluorescence microscopy, we discovered that UvrD self-associates into a tetramer and, unlike Rep, is not recruited to a specific replisome protein despite being found at approximately 80% of replication forks. By deleting *rep* and DNA repair factors *mutS* and *uvrA*, perturbing transcription by mutating RNA polymerase, and antibiotic inhibition; we show that the presence of UvrD at the fork is dependent on its activity. This is likely mediated by the very high frequency of replication blocks due to DNA bound proteins, including RNA polymerase, and DNA damage. UvrD is recruited to sites of nucleoprotein blocks via distinctly different mechanisms to Rep and therefore plays a more important and complementary role than previously realised in ensuring successful DNA replication.

## INTRODUCTION

Replication and transcription of DNA occur simultaneously in *Escherichia coli,* making conflicts between the bacterial replisome, the molecular replication machine comprising in excess of 10 different proteins, and RNA polymerase (RNAP) inevitable (1–4). Collisions between the replication and transcription machineries hinder replication fork progression and cause genome instability (5–14). RNAP can pause, stall and backtrack while actively transcribing, and any immobile RNA polymerases present long-lived barriers to replisomes (15, 16). These factors make RNAP the most common nucleoprotein obstacle for translocating replisomes. DNA is also frequently damaged in normal growth conditions due to exposure to endogenous and exogenous DNA damaging agents. The resulting DNA nicks and mismatches also present significant obstacles for normal replication (17–19).

Several mechanisms exist to reduce conflict between the replisome and obstacles such as transcribing RNA polymerases. For example, transcription elongation and termination factors reduce the number of immobile RNA polymerases on DNA (1, 18, 20–23). Head-on collisions were originally thought to be more harmful (24–26) and therefore more important to resolve; a view supported by the occurrence of highly transcribed genes encoded on the leading strand, presumably to ensure co-directional transcription and replication (25, 27, 28). However, co-directional collisions of the replisome and transcription machinery also impact replication fork progression (3, 18, 29).

Accessory replicative helicases promote fork movement through nucleoprotein barriers, and have been identified in prokaryotes and eukaryotes (1, 12, 13, 30, 31). The Rep and UvrD accessory helicases in *E. coli*, as well as the Pif1 helicase in *Saccharomyces cerevisiae*, perform important roles in helping to clear nucleoprotein complexes (32). More recently, Pif1 has been proposed to be a general displacement helicase for replication bypass protein blocks, a function similar to the Rep and UvrD helicases (33). These helicases can reduce replisome pausing at several types of native transcription complexes observed in living cells (13, 22, 34), and several types of engineered nucleoprotein blocks in biochemical experiments *in vitro* (1, 30).

Rep is the primary accessory helicase in *E. coli*, however, UvrD can partially compensate for the absence of Rep *in vivo* and has been shown to promote replication through stalled transcription elongation complexes (1, 12, 13, 35, 36). UvrD is a superfamily 1 helicase that translocates 3'-5' along ssDNA. The ability of UvrD to compensate partially for the absence of Rep has been attributed to the high degree of homology between these two helicases and the abundance of UvrD inside cells (12, 21, 37–39). The role of UvrD in nucleotide excision repair along with UvrABC proteins as well as in methyl-directed mismatch repair along with the MutHLS proteins is well characterised (40–45). UvrD interacts directly with RNAP but the functional importance of this interaction is unclear (46–48). UvrD has been suggested to promote backtracking of stalled RNAP as a first step in transcription-coupled repair; however, other studies argue against UvrD playing any role in coupling nucleotide excision repair to stalled transcription complexes (47–53).

Here, we combined state-of-the-art live cell imaging using single-molecule microscopy with advanced biochemical activity experiments *in vitro* to elucidate how UvrD resolves replicative blocks. We used super-resolved high-speed single-molecule live cell fluorescence microscopy of mGFP-UvrD in live cells co-expressing a fork marker, DnaQ-mCherry, to reveal that 80% of DnaQ foci colocalise with UvrD. Stepwise photobleaching intensity analysis showed that UvrD operates predominantly as a tetramer *in vivo*, consistent with earlier observations performed *in vitro* (54). Surface plasmon resonance (SPR) measurements of UvrD in combination with each replisome component identified no specific replisome protein interaction partner that would explain replisome-UvrD colocalisation. Deletion of Rep, mismatch repair protein MutS and nucleotide excision repair protein UvrA, and disruption of transcription using either mutation of RNAP or antibiotic inhibition, significantly reduces the number of UvrD or level of observed colocalisation of UvrD with the fork. Coupled with *in vitro* block-dependent differences in UvrD replication promotion efficiency, we conclude that UvrD is located at the *E. coli* replisome due to frequent blocks to replication.

**MATERIAL AND METHODS**

**Strain construction**

All strains used in this study (listed in full in SI Table S1, with associated plasmids and primers in SI Tables S2 and S3) are derivatives of the laboratory wild type strain TB28. Tagging of *dnaQ-mCherry-<kan>* is described in (13). *mGFP-uvrD-<kan>* was amplified from plasmid pAS79 (*uvrD+*) with primers oAS145 and oJGB383 having 50 bp homology to either end of the native *uvrD* locus. The resulting PCR product had homology either side such that recombination with the chromosome would result in integration of the PCR product at the native locus under the control of the native promoter. Prior to integration, all PCR products were treated with DpnI, gel purified, and introduced by electroporation into cells expressing the lambda Red genes from the plasmid pKD46 (55). The recombinants were selected for kanamycin resistance and screened for ampicillin sensitivity. The colonies obtained were verified for integration by PCR and sequencing. The *uvrD* recombinants were verified by PCR amplification using primers oPM319 and oPM320, and sequencing using the primers oJGB417, oJGB418, oPM407, oPM409, oPM411, and oMKG71. Where required, the kanamycin resistance gene was removed by expressing Flp recombinase from the plasmid pCP20 (55) to generate kanamycin sensitive strains carrying the fluorescent protein (FP) fusions. Dual labelled strains were created by introducing the kanamycin tagged FP alleles by standard P1 mediated transduction into single labelled strains carrying the required FP allele after removing the linked kanamycin marker.

**Determination of cell doubling time**

*E. coli* strains were grown overnight in LB medium at 37°C at 200 rpm shaking. The saturated overnight cultures were diluted 100 fold into fresh LB or washed once with 1X 56 salts and diluted 100 fold in fresh 1X 56 salts with 0.2% glucose as the carbon source. Aliquots of 100 µl each of the diluted cultures in fresh media were pipetted into individual wells of 96 well clear flat bottom sterile microplates (Corning). The microplates containing the diluted cultures were incubated in a BMG LABTECH SPECTROstar Nano microplate reader at 37°C and the optical density ($A_{600}$) values were recorded at defined time intervals. The time taken for the optical density values to increase two fold during the exponential growth phase of the culture was taken as the cell doubling time (SI Table S4, Supplementary Figure S1).

**Overexpression and Purification of mGFP-UvrD**

*mGFP-uvrD* was excised from pAS79 using XhoI and SalI and ligated into pET14b cut with XhoI, creating pAS152 encoding histidine-tagged mGFP-UvrD. pAS152 was used to overexpress the mGFP-UvrD fusion in Rosetta™ 2(DE3)/pLysS. The cells were grown in LB medium containing 100 µg/ml ampicillin and 30 µg/ml chloramphenicol at 37°C and 220 RPM to $OD_{600}$ ~ 0.7. The cultures were then equilibrated to room temperature and overexpression was induced by adding 1mM IPTG and incubating at 16°C for 24 hours. Cells from the expressed culture were collected by centrifugation at 4°C and frozen at -80°C. The cells were later thawed on ice and resuspended in buffer A (20mM Tris-Cl pH 8.3, 500 mM NaCl, 5 mM imidazole, 10% glycerol) along with 0.1 mg/ml lysozyme, 1 mM AEBSF, 0.7 µg/ml pepstatin, 0.5 µg/ml leupeptin, 2.5 µg/ml DNase I and 20 µg/ml RNase A and incubated on ice for 30 minutes. Brij-58 was then added to a final concentration of 0.1% v/v and incubated on ice for another 20 minutes. The suspension was sonicated and then clarified by centrifugation at 18,000 RPM for 60 minutes at 4°C in an SS-34 rotor to remove debris. The lysate containing mGFP-UvrD was applied to a 5 ml His-trap FF crude column (GE healthcare) and the protein was purified by chromatography on an imidazole gradient in buffer A. Peak fractions were collected and concentrated using a Vivaspin 500 concentrator (30 kDa MWCO) (Sartorius). Samples were then aliquoted and flash frozen in liquid nitrogen before storage at -80°C. Protein concentration was determined by Bradford's assay.

**Helicase assay**

Assays for unwinding of streptavidin-bound forks were performed using a substrate made by annealing oligonucleotides oPM187 and oPM188. Reactions were performed in 10 µL volumes containing 40 mM HEPES (pH 8); 10 mM DTT; 10 mM magnesium acetate; 2 mM ATP; 0.1 mg ml$^{-1}$ BSA and 1 nM forked DNA substrate. Reactions were carried out as described earlier (13, 56). Briefly, the reaction mixture was pre-incubated at 37°C for five minutes with or without 1 µM streptavidin (Sigma-Aldrich), to which histidine-tagged helicase (as indicated) and biotin (Sigma-Aldrich) were added to 100 µM (acting as a trap for free streptavidin) and incubated at 37°C for another 10 minutes. Reactions were stopped by addition of 2.5 µl of 2.5% SDS, 200 mM EDTA and 10 mg ml$^{-1}$ of proteinase K. Reactions were then analysed by non-denaturing gel electrophoresis on 10% polyacrylamide TBE gels. The quantification of the unwinding and displacement of streptavidin from the fork was carried out as described in (57).

**Microscopy and image analysis**

Imaging was performed on a bespoke dual-colour single-molecule microscope (13). Excitation from Obis 488nm and 561nm 50mW lasers was reduced to 10µm at full width half maximum field in the sample plane producing Slimfield illumination (58). Lasers were digitally modulated to 5ms period to produce alternating laser excitation using National Instruments dynamic I/O module NI 9402. Excitation was coupled into a Zeiss microscope body with the sample mounted on a Mad City Labs nanostage. Images were magnified to 80nm/pixel and imaged using an Andor Ixon 128 emCCD camera. Colour channels were split red/green using a bespoke colour splitter consisting of a dual-pass green/red dichroic mirror centred at long-pass wavelength 560nm and emission filters with 25nm bandwidths centred at 542nm and 594nm.

Cells were imaged on agarose pads suffused with minimal media as described previously (59). Foci were detected and tracked using previously described bespoke MATLAB software (60). In brief, bright candidate foci were detected in each frame by image transformation and thresholding. A 17x17 pixel region of interest (ROI) is drawn around each candidate and subjected to iterative Gaussian masking (61). Foci were accepted if their signal to noise ratio was above 0.4. Foci in successive frames were linked together into trajectories based on nearest distance, provided it was < 5 pixels. Linked foci were accepted as "tracks" nominally if they persist for at least 4 consecutive image frames.

Characteristic intensity of mGFP or mCherry was determined from the distribution of foci intensity values towards the end of the photobleach confirmed by overtracking foci beyond their bleaching to generate individual photobleach steps of the characteristic intensity (Supplementary Figure S5). This intensity was used to determine the stoichiometry of foci by fitting the first 3 intensity points with a straight line and dividing the intercept by this characteristic intensity. The number of peaks in the Gaussian fits to UvrD was set by running a peak fitting algorithm over the wild type distribution. This number of Gaussians was then used for mutant distributions unless two or more of the Gaussians converged on the same/similar peak value, in which case they were removed. For DnaQ, two peaks were fit as used previously (62).

GFP and mCherry images were aligned based on the peak of the 2D cross correlation between their respective brightfield images. Colocalisation between foci and the probability of random colocalisation was determined as described previously (63).

The microscopic apparent diffusion coefficients ($D$) were determined by fitting a straight line to the first three mean squared displacements (MSD) values constrained through the equivalent localization precision MSD as determined from the intensity (64). $D$ distributions were fit by three Gamma distributions as described previously (65). Dwell time was calculated as the number of frames that each trajectory was colocalised with the fork position, as determined by the DnaQ foci detected at time zero.

**Surface plasmon resonance (SPR)**

SPR was performed at 25°C on a BIAcore 2000 instrument as in (12). Immobilisation of *E. coli* UvrD and Rep and *Bacillus subtilis* PcrA was performed onto streptavidin-coated SA sensor chips whilst the indicated concentrations of Tau were passed over in 10 mM HEPES pH 7.4, 3 mM EDTA, 150 mM NaCl, 10 mM MgCl2 and 0.005% Tween 20 at 20 µl min-1. This buffer differed from that used in the DNA replication assays to minimise non-specific interactions with the surface-immobilised streptavidin.

**Replication assay**

Replication assays were carried out using pPM872 as the template for RNAP blocks (1) or pPM594 (12) for EcoRI E111G blocks. Plasmid pPM872 contains $P_{lacUV5\ 52C}$, a strong promoter in which the first 52 nucleotides of the transcript lack cytosine residues and are then followed by four consecutive cytosines. This enables transcription of $P_{lacUV5\ 52C}$ to be stalled by the omission of CTP. Assays were performed in 40 mM HEPES (pH 8); 10 mM DTT; 10 mM magnesium acetate; 2 mM ATP; 0.2 mM GTP, 0.2 mM UTP; 0.04 mM dNTPs; and 0.1 mg/ml BSA. Reactions (15 µl) contained 2 nM plasmid template, 50 nM DNA polymerase III $\alpha\epsilon\theta$ complex, 25 nM $\tau$ clamp loader complex, 160 nM DnaB and DnaC monomers, 1 µM SSB, 80 nM β clamp, 30 nM HU, 200 nM DnaG, 300 nM DnaA. Helicases were added as indicated. Rep and UvrD used at 200 nM. *E. coli* RNAP holoenzyme from NEB (1 U/µl) was used at 1/3 dilution. Dilution was determined empirically to match RNAP replication inhibition levels from (1). EcoRI E111G was purified as in (66) and used at 200 nM. Reactions were assembled on ice and initiated by addition of DnaA and incubation for 4 min at 37°C, followed by addition of 60 units SmaI (Promega) plus 0.4 MBq [$\alpha^{32}$P] dCTP (222 TBq/mmol). Standard reactions with the helicase present in the initial protein mix were carried out at 37°C for 1 minute and then terminated by addition of 1 µl of 0.5 M EDTA. Delayed helicase addition was carried out by adding Rep/UvrD after 1 minute and incubation for a further minute at 37°C before termination with EDTA. Ethanol precipitated replication products were analysed by denaturing agarose gel electrophoresis (0.7% agarose in 2 mM EDTA 30 mM NaOH for 400 volt hours, standard run was 16 hours at 25 V), phosphorimaging and autoradiography. 5'-labelled HindIII- digested lambda DNA was used as a marker. Gels were quantified using Quantity One® (Bio-Rad) software. Full-length plasmid replication products were quantified as a proportion of summed blocked products and normalised for length (for EcoRI E111G block), or total lane signal (RNAP block).

# RESULTS

## UvrD is present at the majority of replication forks

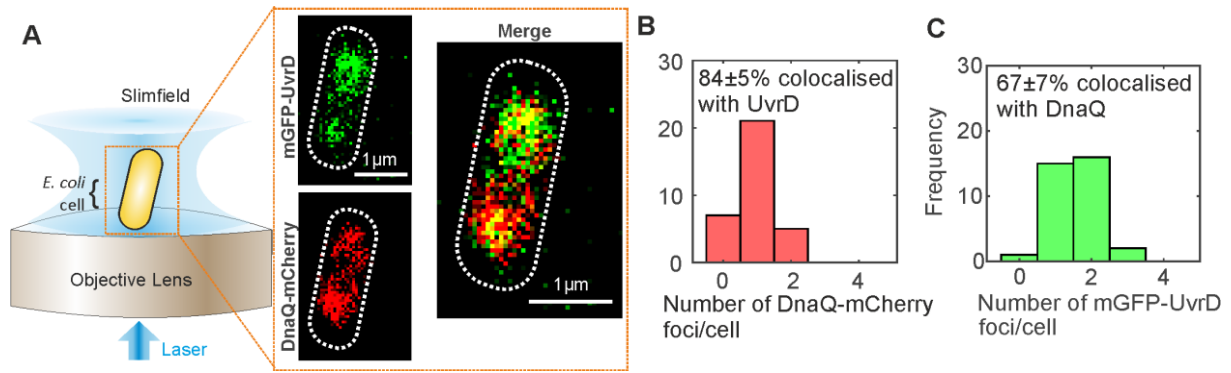

**Figure 1: Super-resolved single-molecule light microscopy.** (A). Slimfield microscope schematic and Slimfield micrographs of mGFP-UvrD, DnaQ-mCherry. (B) and (C) Histogram showing the number of DnaQ and UvrD foci detected per cell respectively, SD errors.

We investigated the role of UvrD in DNA replication using super-resolved single-molecule fluorescence microscopy to track dynamic patterns of fluorescently labelled UvrD localization relative to the replication fork in live cells (Figure 1A). We performed dual-colour Slimfield (58) to assess the degree of colocalisation between UvrD and the replisome, using a similar imaging protocol as previously for Rep (13). We employed the same DnaQ-mCherry fork marker as before along with a genomically integrated mGFP-UvrD fluorescent protein fusion construct to report on the localization of UvrD. DnaQ encodes the epsilon subunit, a core component of the DNA polymerase. These strains phenocopied the wild type strain in growth and plasmid loss assays (Supplementary Figures S1,S2, Tables S1-4). The helicase activity of mGFP-UvrD was comparable to the wild type enzyme in an *in vitro* unwinding assay (Supplementary Figure S3). Under Slimfield, we observed 1-2 DnaQ foci as in the previous study for Rep, corresponding to the two moving replication forks which appear as single foci near the start of replication when they are separated by less than the diffraction limit of our microscope of 200-300 nm (Figure 1A,B Supplementary Figure S4). We also observed 1-2 UvrD foci per cell, of which 67±7% (±SEM) were colocalised with DnaQ foci and 84±5% of DnaQ foci were colocalised with UvrD (Figure 1B and C). We calculated the probability of random overlap between these foci, by modelling the nearest-neighbour foci separation as a Poisson distribution (63), to be 20%, implying that, similarly to the Rep helicase (13), UvrD is present at the majority of replication forks.

## UvrD is a tetramer in live bacterial cells

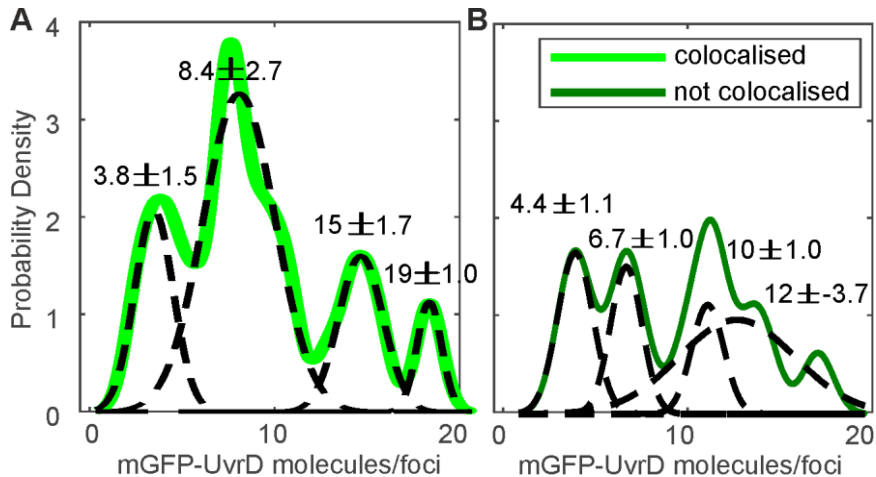

**Figure 2 UvrD stoichiometry.** The distribution of UvrD foci stoichiometry rendered as a kernel density estimate for foci colocalised (A) and not colocalised (B) with DnaQ foci. Distributions were fitted with multiple Gaussian peaks (black dotted lines) with the peak value ± half width at half maximum indicated above each peak.

Slimfield also enabled the stoichiometry of these foci to be determined, utilizing a method involving step-wise photobleaching analysis (62). Here, we determined the brightness of a single-molecule of mGFP or mCherry fluorescent protein under our Slimfield imaging conditions (Supplementary Figure S5), and used these values to normalize the initial brightness of all tracked foci as an estimate for the apparent stoichiometry in terms of number of fluorescently labelled molecules present. We found that DnaQ had the same range of 2-6 molecules per focus (Supplementary Figure S6) as we have found previously, corresponding to 2-3 polymerases per fork (13, 62). The distribution of the apparent stoichiometry of UvrD foci which are colocalised with DnaQ had a distinct lowest-order peak corresponding to approximately four molecules, with subsequent peaks at clear multiples of four (Figure 2A). UvrD foci that are not colocalised with DnaQ also contained a distinct lowest-order peak at four but subsequent peaks were less clearly tetrameric. UvrD has been shown to self-associate into tetramers *in vitro* (54) with DNA unwinding experiments *in vitro* also suggesting UvrD must function as at least a dimer (67) or a trimer (68). To our knowledge, these are the first UvrD stoichiometry measurements *in vivo* so we propose that UvrD forms a tetramer *in vivo* with multiple tetramers often at the replication fork. It should be noted that we often observe only a single DnaQ focus corresponding to two replication forks which are separated by less than the diffraction limit of our microscope (Supplementary Figure S6) such that some multiple tetramer UvrD foci correspond to two replication forks. The less distinct peaks in the UvrD foci that are not colocalised with DnaQ may indicate that other UvrD species (monomers, dimers or trimers) are also present.

We were also able to determine the total cell copy number of UvrD using a method which determined the total contribution of mGFP-UvrD fluorescence to the observed intensity in whole cells (64). We found approximately 800 molecules of UvrD per cell (Supplementary Figure S7), similar to that which we reported previously for Rep (13). This is a much lower level of UvrD than was reported using transcription rate measurements (69) or high-throughput proteomics (70) but is comparable to recent estimates using ribosome profiling (71).

**UvrD has no interaction partner at the fork**

Our earlier findings concerning the accessory helicase Rep indicated that colocalisation with the fork and recruitment to the replisome were mediated by specific interaction with the replicative helicase DnaB (72), with Rep and DnaB both exhibiting hexameric stoichiometry (13, 62). We sought to determine whether UvrD is also recruited by interacting with a specific replisome component using SPR on purified UvrD and each replisome component (Supplementary Figure S8). UvrD exhibited no direct interaction with primase, SSB, β sliding clamp, the DNA polymerase III αε, χψ or γ complexes; or θ, δ, δ′, χ, and γ subunits (12). We did initially identify a putative UvrD-Tau interaction. However, as a control we also tested the *B. subtilis* helicase PcrA, and found this also interacted with Tau to a similar extent. We therefore conclude that the Tau interaction with UvrD is likely non-specific and that, unlike Rep, UvrD likely does not have a direct interaction partner at the replisome from the suite of replisome proteins tested.

**UvrD presence at the replication fork is mediated by specific blocks to replication**

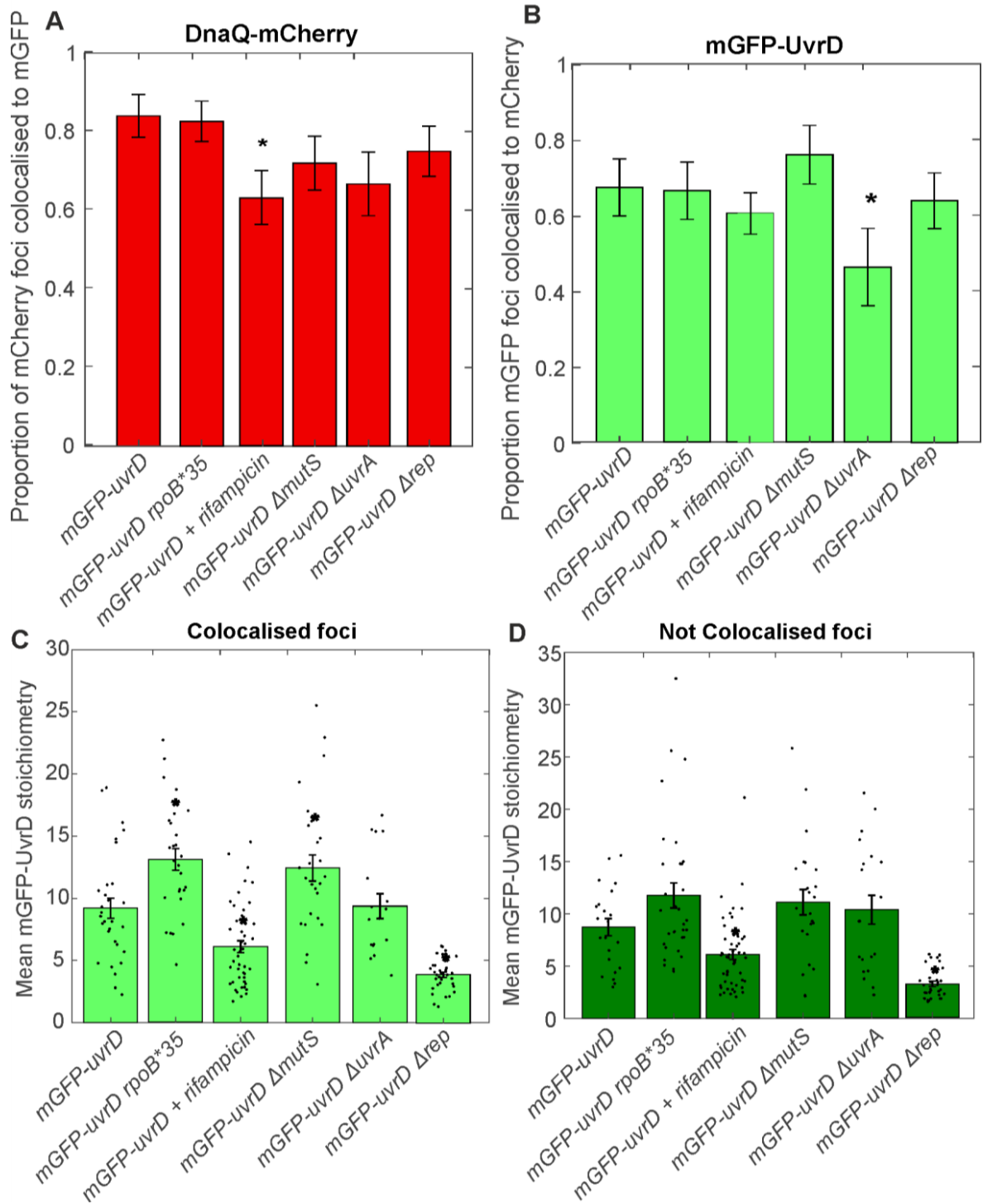

**Figure 3: UvrD perturbations.** A. and B. The mean proportion of colocalised DnaQ and UvrD foci per cell respectively. Error bars represent standard error in the mean and statistical significance by Student's t-test p<0.05 indicated by *. C. and D. Jitter plots of colocalised and not colocalised UvrD stoichiometry respectively. Bars show mean UvrD stoichiometry with errorbars showing the standard error in the mean and statistical significance by Student's t-tests p<0.05 indicated by *.

Since our SPR observations indicated that UvrD does not interact directly and specifically with any core replisome protein, we hypothesized that the association of UvrD to the replication fork might be dependent on its activity in resolving nucleoprotein blocks. To test this, we systematically impaired key DNA replication repair processes in which UvrD is implicated, as well as perturbing RNAP replicative blocks by disruption of transcription through both genetic mutation and antibiotic inhibition (Figure 3). We perturbed UvrD's activity in resolving DNA damage at replication blocks by deleting *mutS* or *uvrA*, rendering the strains defective in UvrD-associated mismatch repair or nucleotide excision repair respectively (42, 73). We impaired UvrD's activity in removing transcriptional blocks to replication (48) by introducing the *rpoB*35* mutation, which destabilizes open complexes formed during initiation of transcription (74, 75); and by using the antibiotic rifampicin, which blocks elongation of RNA transcripts (76). All of these strains were healthy under normal growth conditions (Supplementary Figure S1, Table S4) and displayed the same phenotype for DnaQ foci number and stoichiometry within measurement error (Supplementary Figure S6).

For the transcriptional block perturbations, the *rpoB*35* mutant dual labelled strain exhibited no difference in the proportion of colocalised DnaQ or UvrD foci (Figure 3A and B). This mutant and all others tested exhibited the same predominant tetrameric stoichiometry trends as wild type (Supplementary Figure 9), with clear peaks at tetramer intervals in the colocalised stoichiometry distribution and less clear but still largely tetrameric intervals in the stoichiometry distribution for UvrD foci that are not colocalised with DnaQ. This provides further evidence that UvrD operates predominantly as a tetramer *in vivo*. We compared the mean stoichiometry for each mutant to that of wild type (Figure 3C and D). Unexpectedly, the *rpoB*35* mutant produced a slight but statistically significant increase in mean colocalised stoichiometry, corresponding to approximately one extra UvrD tetramer at the replication fork. This finding may indicate that *rpoB*35* does not alter frequency of transcriptional blocks to replication or that the level of instability conferred by this mutation is not sufficient to clear RNAP blocks to alter UvrD localisation significantly at the fork. Alternatively, it could mean that *rpoB*35* does not affect the formation of the types of transcriptional block that are acted upon by UvrD. Rifampicin treatment produced a clearer response, reducing the number of DnaQ foci colocalised with UvrD by ~20% (Figure 3A) and the mean stoichiometry of colocalised and not colocalised

UvrD foci by around one UvrD tetramer (Figure 3C and D). Single-molecule observations of RNAP have shown it to be significantly more mobile under rifampicin treatment (77), implying fewer RNAPs bound to DNA and fewer blocks to replication. Our results are consistent with fewer transcriptional blocks to replication reducing the amount of UvrD recruited to the replisome and support the hypothesis that UvrD is recruited by these replicative blocks.

By deleting *mutS* or *uvrA* we removed UvrD's role in resolving mismatch repair and nucleotide excision repair respectively. Deleting *mutS* surprisingly resulted in no change in UvrD colocalisation with DnaQ but did increase the mean UvrD stoichiometry colocalised with DnaQ by around one tetramer (Figure 3A). In the ∆*uvrA* mutant this resulted in a clear drop of ~25% in UvrD colocalisation with DnaQ (Figure 3B), consistent with the hypothesis that UvrD association with the fork is dependent on actively resolving DNA damage induced blocks to replication. The minimal response in the ∆*mutS* strain may also suggest that UvrD at the fork is more involved in nucleotide excision repair than mismatch repair.

UvrD has been shown to compensate for Rep in *rep* null mutants (12). We also tested what effect this mutation would have in our single-molecule assay. Surprisingly, *rep* deletion resulted in a steep drop in UvrD copy number (Supplementary Figure S7). The degree of colocalisation between UvrD and DnaQ remained unchanged compared to wild type (Figure 3A and B) but the stoichiometry of foci was reduced to a single UvrD tetramer (Figure 3C and D and Supplementary Figure 9). The drop in UvrD copy number may reflect complex regulatory shifts in response to losing such a key repair protein in Rep, although it is interesting that even at low copy numbers, a UvrD tetramer remains recruited to the fork. Double knockouts, deficient in both Rep and UvrD, are lethal (12) so this minimal helicase presence must be the limit required for survival.

**UvrD dynamically associates with the replication fork dependent on repair processing**

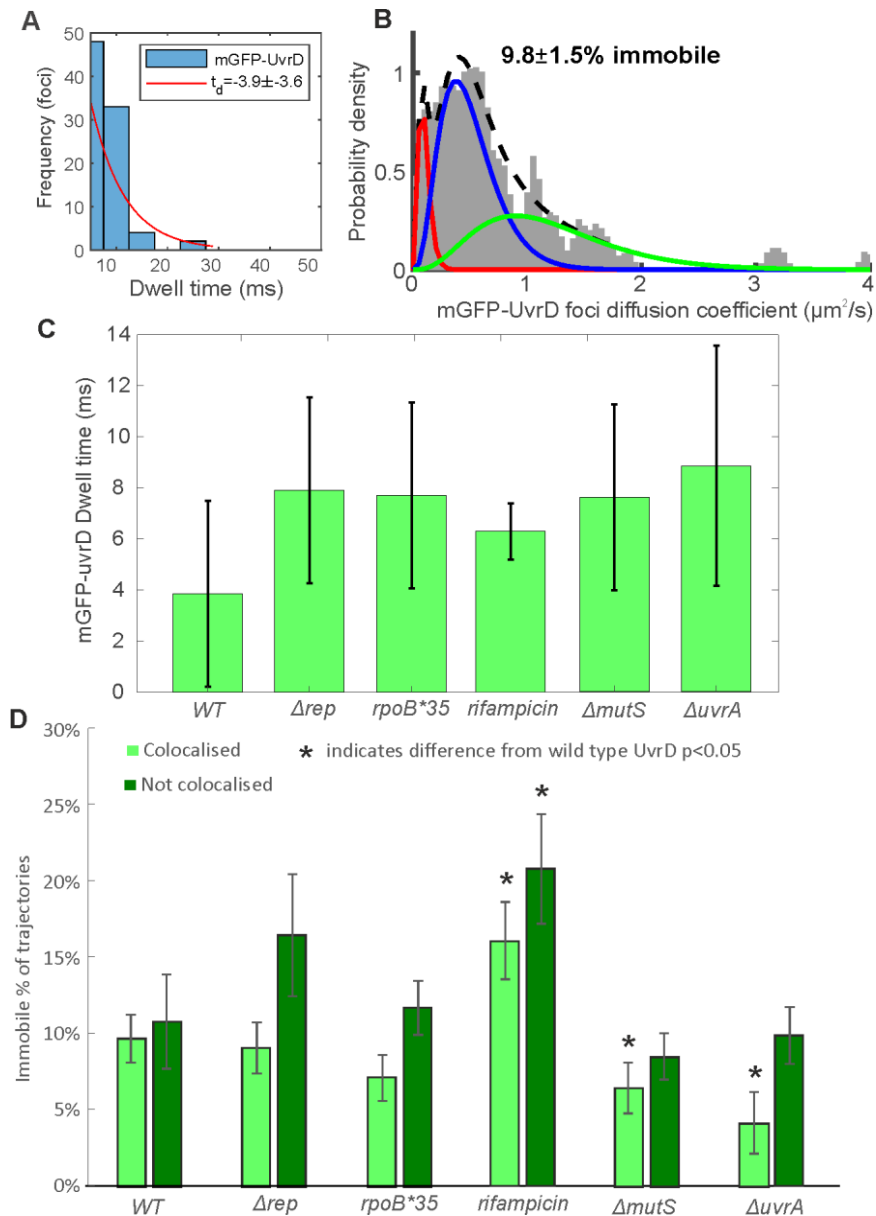

**Figure 4: UvrD dynamics.** A. Distribution of time UvrD and DnaQ foci were colocalised (blue) fit with an exponential (red) to yield a characteristic dwell time. B. Distribution of microscopic diffusion coefficients of wild type UvrD foci (grey) fitted with a three state gamma distribution model containing a relatively immobile population with $D = 0.1$ µm$^2$/s (red), a transiently immobile population, $D = 0.5$ µm$^2$/s (blue) and a mobile population, D=1.2 µm$^2$/s (green). C. The dwell times for each UvrD strain. Error bars represent the 95% confidence intervals on the fit. D. The proportion of UvrD foci in the immobile population for each UvrD strain colocalised or not colocalised with DnaQ foci location. Statistically significant differences p<0.05 from wild type indicated by *.

By tracking UvrD foci as a function of time using rapid millisecond timescale sampling in Slimfield imaging, we were also able to characterize its dynamics in wild type and repair impaired cells. We measured the number of frames UvrD foci were colocalised with DnaQ foci positions, the dwell time, and calculated the apparent microscopic diffusion coefficient of individual foci (Figure 4 and Supplementary Figure 10). The UvrD dwell time at the fork for the wild type strain was measured to be approximately 4 ms, indicating a relatively dynamic association of UvrD at the replication fork. The apparent microscopic diffusion coefficients ($D$) were best fit with a three parameter Gamma model, containing a relatively immobile population with $D = 0.1$ µm$^2$/s, a transiently immobile population, $D = 0.5$ µm$^2$/s and a mobile population, D=1.2 µm$^2$/s, similar to Rep (13). In wild type, we found approximately 10% of UvrD foci were immobile, whether colocalised with the DnaQ fork location or not, with the rest of the foci split between the other two populations.

We sought to use these molecular measurements to determine the effect of perturbing UvrD-mediated DNA replication repair and block resolution on UvrD dynamics. All deletions resulted in a marginal though statistically negligible increase in dwell time to between 6-9 ms. The Δ*rep* and *rpoB\*35* mutations resulted in no change in mobility from wild type. Rifampicin treatment however, resulted in an increase in the proportion of immobile foci both at and away from the fork. This finding appears counterintuitive as rifampicin treatment results in more mobile RNAP (77), however it is unclear what effect rifampicin has on the population of RNAP molecules that remain bound to DNA. These results may suggest that such blocks provide an increased barrier to replication, decreasing mobility of UvrD as it attempts to deal with these blocks. In the Δ*mutS* and Δ*uvrA* strains, the proportion of immobile foci decreased at the fork, again consistent with the hypothesis that UvrD association with the fork is dependent on block resolution activity. An alternative explanation for these results is that UvrD is less mobile when resolving DNA damage blocks to replication and more mobile when resolving transcriptional blocks, as reducing or eliminating UvrD's role in these processes decreased or increased the immobile fraction in DNA repair and transcription impaired cells respectively (Figure 4).

### *In vitro* experiments confirm block-dependent UvrD recruitment

To further investigate the role of blocks in UvrD recruitment to the replisome, we performed a stalled replication assay *in vitro,* which allowed us to specifically vary the block and the timing of UvrD recruitment i.e. before or after the replisome had encountered the block. *E. coli* replisomes can be reconstituted on plasmid DNA templates containing *oriC* and engineered stall sites. UvrD can then be included from the start or added after the replisome has encountered the block. Blocked fork resolution was then probed by examining the replication products by denaturing gel electrophoresis.

We first used an *oriC*-containing plasmid which contained a specific RNAP stall site ($P_{lacUV5\ 52C}$) to facilitate formation of a stalled transcription elongation complex via nucleotide deprivation. Stalled RNAP acts as a barrier to the replisome in both head-on and co-directional orientations and prevents the formation of full-length replication products (1). Arresting the replisome at this RNAP block resulted in replication products consisting of four truncated leading strands. The strand sizes matched those expected for replisomes moving clockwise or counter-clockwise from *oriC* and encountering $P_{lacUV5\ 52C}$ or the promoters within the *ColE1* plasmid origin of replication (Figure 5A, lane 2 and Figure 5C) (1).

Inclusion of UvrD promoted movement of replisomes through stalled RNAP and resulted in a decrease in all four truncated leading strand products and a concomitant increase in production of full-length leading strands (Figure 5A, lane 5 and 5E) (1). This was the case when UvrD was included in reactions before the initiation of replication. Addition of UvrD once the replisome had already encountered the block gave variable results that ranged from an inhibitory to a stronger effect, although differences were not statistically significant at $p<0.05$ (Figure 5A, compare lane 5 with 6, Figure 5E). Similar results were also obtained for Rep (Figure 5A, compare lane 3 with 4, Figure 5E). In the absence of replication neither Rep nor UvrD are capable of pushing stalled RNAP from the $P_{lacUV5\ 52C}$ stall site (1).

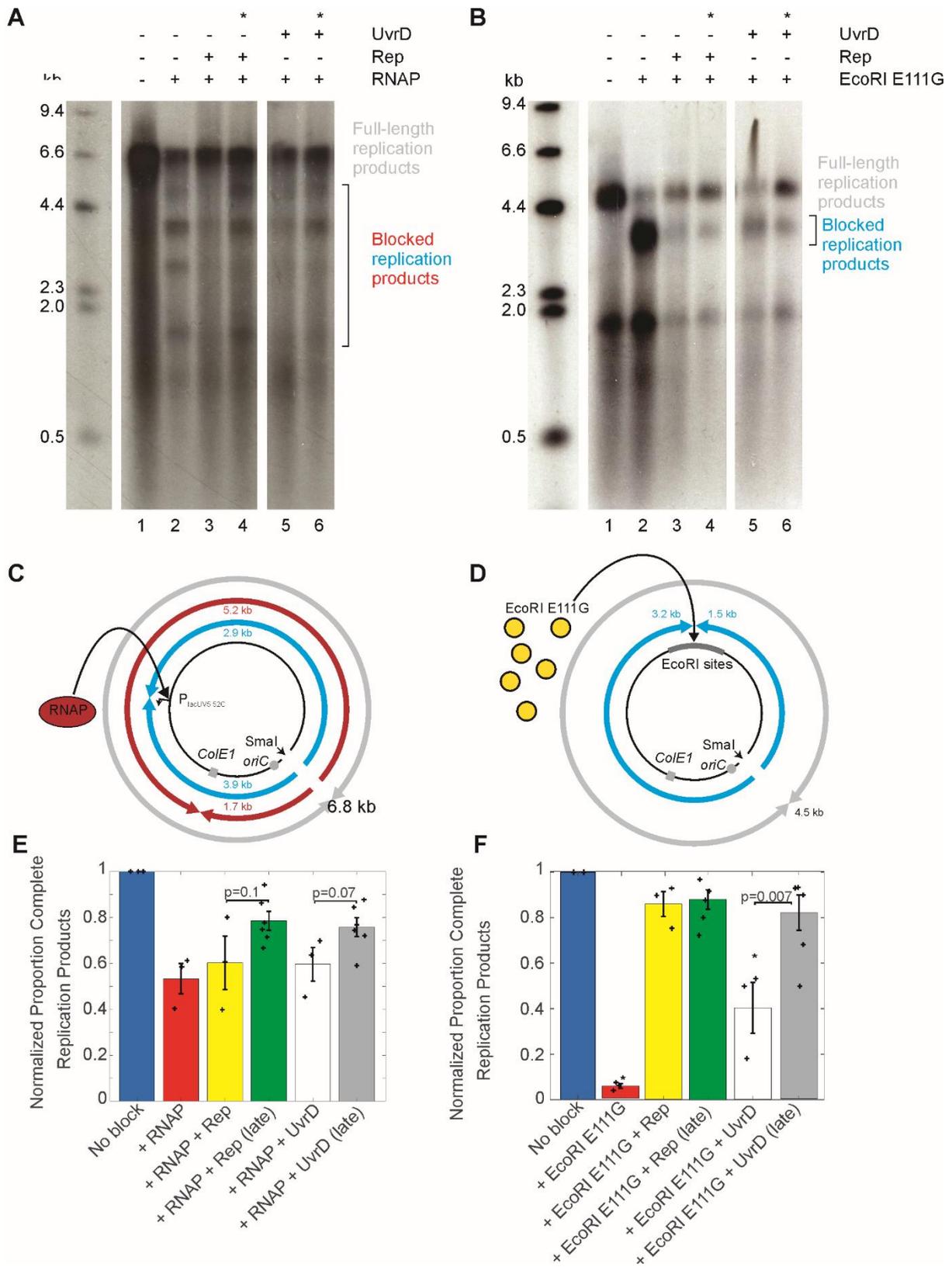

**Figure 5: Helicase pre-association affects the promotion of replication fork movement through different nucleoprotein blocks** A. Denaturing agarose gel of replication products formed from pPM872 (1) in the absence and presence of RNAP block, Rep and UvrD added pre- and post-collision (*) as indicated. B. Denaturing agarose gel of replication products

formed from pPM594 (12) in the absence and presence of EcoRI E111G block, Rep and UvrD added pre- and post-collision (*) as indicated. Truncated leading strand products formed by replisomes originating from *oriC* and colliding with RNAP (C) and EcoRI E111G (D). Quantification of lanes in (A) and (B) from three technical replicates shown in (E) and (F), with standard error in the mean indicated and each replicate result shown as a cross.

Since UvrD has been shown to interact with RNAP (46–48), we tested whether UvrD-block pre-association is a requirement for its promotion of replication. EcoRI E111G is a mutant restriction enzyme that binds to its recognition sequence but has reduced cleavage activity (78). EcoRI E111G can thus act as an engineered replication fork block that can be overcome by Rep or UvrD helicase activity (12). We tested whether UvrD helicase could promote replication through this barrier when added before or after initiation of replication on a plasmid template with an array of eight EcoRI sites (Figure 5B and D). Adding UvrD post-collision did not impede its promotion of replication; in fact the resolution of these collisions appeared to be more efficient because the proportion of full-length replication products increased (Figure 5B, compare lane 5 and 6, 5F). Quantification of full-length replication products as a proportion of reaction products showed that late UvrD addition increased full-length replication products to 84% (n=6) from 41% (n=3) when UvrD is present before fork collision. This contrasts with the RNAP block results where the UvrD-RNAP interaction (46–48) might be responsible for localising UvrD to the block and facilitating resolution, resulting in no difference when present pre- or post-collision. Similarly, Rep associates directly with the replication fork, also resulting in no difference when present pre- or post-collision, for both blocks. These results suggest that UvrD might recognize and bind stalled replication forks at blocks with a different mechanism to Rep. Combining these data with the lack of a specific replisome partner and reductions in colocalisation between UvrD and DnaQ fork marker when replication blocks are perturbed *in vivo*, leads us to propose that UvrD is present at the replication fork due to the high frequency of blocks to replication.

**DISCUSSION**

Here, we show that as with the Rep helicase, the majority of replisomes are also associated with UvrD but unlike Rep, UvrD has no interaction partner in the replisome. UvrD instead associates with the replisome through the frequent blocks to replication encountered by the fork. Similarly to Rep, *in vitro* evidence suggests UvrD functions poorly as a helicase when monomeric (79). We showed previously that Rep functions as a hexamer *in vivo* but this oligomerisation is due to interactions with hexameric DnaB (13). Again UvrD functions differently and appears to self-associate into tetramers, as evidenced here by tetramer periodicity at and away from the fork (Figure 2 and Supplementary Figure 9) and previous ultracentrifugation experiments *in vitro* (54). It is possible that UvrD may not have evolved to fulfil exactly the same function as Rep; as discussed previously, UvrD is involved in a range of DNA repair processes, including nucleotide excision and mismatch repair. Also, it is important to note that it is not uncommon in general in many enzyme-catalyzed reactions that if one enzyme is depleted then another will complement for its function, but with a reduced catalytic efficiency. On initial inspection, this might appear to be the case for Rep and UvrD, in light of the lower efficiency of block resolution for UvrD compared to Rep, noting that deletion of Rep has a profound impact on replication whereas lack of UvrD does not. However, there is a nuance here in that what we have discovered in our present work is that Rep and UvrD have distinctly different modes of recruitment at a replication block. These differing mechanisms of replication promotion may suggest a model in which these accessory helicases have evolved to fulfil the same overall function (*i.e.* that of resolving a block to DNA replication) but in different ways. Even with different block removal efficiency, such redundancy may confer an increased level of robustness, helping to ensure that the vital process of DNA replication is carried out correctly by the cell.

We used single-molecule microscopy to show that a high proportion of forks, 80%, contained UvrD *in vivo.* Although the raw imaging data of our Slimfield microscopy is still subject to the optical resolution; the localization tracking algorithms applied can pinpoint fluorescently labelled UvrD to approximately 40 nm lateral precision, smaller than the 50 nm diameter of the replisome machinery itself (62). Thus, our measurements for colocalisation between UvrD and DnaQ are likely to be good indicators of a real interaction between the replication fork and UvrD. Importantly, the degree of colocalisation we observed is much higher than expectations of random overlap of the optical images of DnaQ-mCherry and mGFP-UvrD.

Through SPR, however, we could not find any interaction partner at the replisome apart from Tau. As a control, we tested Tau's interaction with a related helicase from a different species, PcrA from *B. subtilis*, and also found a positive for interaction. Thus, we conclude that the interaction of UvrD with Tau is likely non-specific and UvrD recruitment to the replisome is mediated by blocks, rather than an interaction partner at the replisome.

We probed UvrD's function at the replication fork by perturbing its known DNA repair processing functions, specifically by knocking out *mutS* and *uvrA*, thus blocking mismatch and nucleotide excision repair respectively. We also introduced a mutant RNAP with *rpoB*35* and treated the wild type with rifampicin to reduce the occurrence of transcriptional blocks to translocating replication machinery. Deleting *uvrA* and rifampicin treatment produced clear results, reduced fork colocalization, but deleting *mutS* and introducing the *rpoB*35* mutation did not. However these latter perturbations did result in an identical phenotype of one extra UvrD tetramer per fork. The *rpoB*35* mutation improves cell viability to increased DNA lesions (74). This intriguing link between mutations requires further study, possibly of these dual labelled strains in response to increased DNA lesions.

The only function of UvrD we did not probe directly here was its role in RecA-mediated recruitment. We believe the results from a *recA* knockdown would be difficult to interpret due to its pleiotropic effects. RecA is the principal recombinase that induces the complex SOS response resulting in the induction of UvrD expression (80, 81). Furthermore, UvrD removes RecA from RecA-DNA complexes, thus functioning as an antirecombinase (82–84). Dissecting these functions of UvrD from other effects that a *recA* mutation may have on the cell would be difficult. Applying the perturbations in combination, i.e. double knockdowns or knockdowns plus rifampicin, would likely lead to confusing results since too many deleterious perturbations may be lethal or change normal cell function too radically to be biologically relevant in the intended way. For the suite of mutant strains investigated we found small but significant changes in colocalisation of UvrD with the replication fork, implying that UvrD is recruited to the replication fork by the frequent blocks to replication it encounters. For example, the UvrD-RNAP interaction could localise UvrD to the vicinity of active transcription machinery. Since RNA polymerase is the most common replication block, a secondary consequence would be UvrD-fork co-localisation. This is in contrast to the Rep accessory helicase which has a specific interaction partner in the replisome of the replicative helicase DnaB (72). Recent biochemical findings *in vitro* show that blocks to replication occur very frequently (1, 19, 85), potentially accounting for our finding that UvrD colocalises with the fork more frequently than Rep, despite Rep having a specific interaction partner at the fork.

We have shown that replication forks stalled at transcriptional blocks can be rescued by addition of UvrD, prior or post-stalling. If UvrD is to help perform this function then it must be present during the resolution of collision between the replication and transcription machinery, however, our estimate for dwell time at the fork of just a few milliseconds suggests a non-negligible dissociation constant; UvrD is not processive *per se* but rather undergoes rapid turnover at the fork such that there is a high likelihood for UvrD being present at any given time, but not necessarily the same set of molecules. However, using a defective restriction enzyme EcoRI E111G block produced a different effect, with late UvrD addition speeding up block resolution. This restriction enzyme block is artificial and less biologically relevant but may indicate that the known specific UvrD interaction with RNAP is important. Improved replication resolution with late addition of UvrD is intriguing, implying increased affinity for or activity at pre-stalled replication forks compared to UvrD being present as the fork stalls. Complete understanding requires further study of replication block resolution in different conditions.

Our results suggest a model where UvrD diffuses and transiently interacts with DNA at random or through specific interactions such as with RNAP or a specific conformation of the blocked replication fork. At the DNA UvrD is predominantly tetrameric, either pre-assembling in solution or on the DNA itself. The frequent sampling leads to UvrD often being in the vicinity of the replisome, allowing UvrD to resolve the frequent blocks to replication. It is also possible that certain barriers to replication may be harder for Rep to resolve, and such barriers can possibly be more efficiently removed by UvrD, however, it is still likely to be the case that whatever they are they do not directly challenge the ability of the cell to replicate the genome in light of *uvrD* deletion still resulting in viable cells. Indeed, UvrD shares a high degree of structural similarity with Rep, although the two proteins differ significantly in their amino acid sequences (30, 86). While Rep is recruited to the fork through specific interaction with DnaB, no such interaction could be identified for UvrD. Stochastic impairment of the ability of Rep to interact with DnaB may necessitate block processing at the fork by UvrD. In summary, our results highlight the contribution of UvrD action to facilitating accurate genome duplication at active replisomes.

**Acknowledgments**

We thank Maria Chechik for help with mGFP-UvrD purification, Jamieson Howard for help with helicase assays. We thank the Molecular Interactions Laboratory in the Bioscience Technology Facility at the University of York for technical assistance and support.

**Funding**

Supported by BBSRC via grants BB/N006453/1 (M.L., P.M.), BB/R001235/1 (M.L., (M.L.,P.M.,A.S.), BB/N014863/1 (M.H.).

# Supplementary Information

## Tetrameric UvrD helicase is located at the *E. coli* replisome due to frequent replication blocks

Adam J. M Wollman[1,2,3,], Aisha H. Syeda[1,2], Andrew Leech[4], Colin Guy[5], Peter McGlynn[2], Michelle Hawkins[2] and Mark C. Leake[1,2]

The authors wish it to be known that, in their opinion, the first two authors should be regarded as joint First Authors

[1] Department of Physics, University of York, York YO10 5DD, United Kingdom.

[2] Department of Biology, University of York, York YO10 5DD, United Kingdom.

[3] Current address: Biosciences Institute, Newcastle University, NE1 7RU, United Kingdom.

* To whom correspondence should be addressed. To whom correspondence should be addressed. Tel: +44 (0)1904322697. Email: mark.leake@york.ac.uk

Present Address: Departments of Physics and Biology, University of York, York YO10 5DD, United Kingdom

## Supplementary Tables
## Table S1 Strains used in this study

| Strain | Relevant genotype | Source or derivation |
|---|---|---|
| BW25113 derivatives: | | |
| BW25113 | *rrnB ΔlacZ4787 hsdR514 Δ(araBAD)567 Δ(rhaBAD)568 rph-1* | (1) |
| JW2703 | *ΔmutS::<kan>* | (1) |
| JW3786 | *ΔuvrD::<kan>* | (1) |
| JW4019 | *ΔuvrA::<kan>* | (1) |
| MG1655 derivatives: | | |
| TB12 | *ΔlacIZYA::<kan>* | |
| TB28 | *ΔlacIZYA* | (2, 3) |
| TB28 derivatives | | |
| N6577 | *ΔlacIZYA Δrep::cat* | (4) |
| N6661 | *ΔlacIZYA::<> Δrep::dhfr* / pAM407 (*uvrD+ lacZ+*, a pRC7 derivative) | (4) |
| N6661 | *ΔlacIZYA::<> Δrep::dhfr* / pAM407 (*uvrD+ lacZ+*, a pRC7 derivative) | (4) |
| N6644 | *ΔlacIZYA::<> uvrD::dhfr Δrep::cat* / pAM407 (*uvrD+ lacZ+*, a pRC7 derivative) | (4) |
| JGB166 | *ΔuvrD* / pKD46 | JW3786 transformed with pKD46 |
| AS97 | TB28/pKD46 | (5) |
| AS446 | *dnaQ-mCherry-<kan>* | |
| AS448 | *dnaQ-mCherry-<>* | (5) |
| AS512 | *mGFP-uvrD-<kan>* | *mGFP-uvrD-<kan>* recombineered into JGB166 |
| AS517 | *ΔlacIZYA::<> Δrep::cat mGFP-uvrD-<kan>* / pAM407 (*uvrD+ lacZ+*, a pRC7 derivative) | N6661 X P1(AS512) |
| AS564 | *dnaQ-mCherry-<kan> rpoB*35* | PM486 X P1(AS446) |
| AS571 | *dnaQ-mCherry-<> rpoB*35* | AS564 to km$^S$ with pCP20 |
| AS580 | *dnaQ-mCherry-<> mGFP-uvrD-<kan>* | AS448 X P1(AS512) |
| AS582 | *dnaQ-mCherry-<> mGFP-uvrD-<kan> rpoB*35* | AS571 X P1(AS512) |
| AS590 | *dnaQ-mCherry-<> mGFP-uvrD-<kan> Δrep::cat* | AS580 X P1(N6577) |
| AS650 | *dnaQ-mCherry-<> mGFP-uvrD-<>* | AS580 to km$^S$ with pCP20 |
| AS656 | *dnaQ-mCherry-<> mGFP-uvrD-<> ΔuvrA::<kan>* | AS650 X P1(JW4019) |
| AS658 | *dnaQ-mCherry-<> mGFP-uvrD-<> ΔmutS::<kan>* | AS650 X P1(JW2703) |

**Table S2 Plasmids used in this study**

| Plasmid | Description | Antibiotic | Reference |
|---------|-------------|------------|-----------|
| pCP20 | Yeast Flp recominase expression plasmid | Amp | (6) |
| pKD46 | λ Red recombinase expression plasmid | Amp | (6) |
| pAS79 | pUC18 *mGFP-uvrD-<kan>* | Amp, kan | This study |
| pAS152 | pET14b-mGFP uvrD | Amp | This study |

**Table S3 Primers used in this study**

| Primer | Sequence (5' - 3') |
|---|---|
| oJGB383 | TGAATGATTTTTTAGGCCAAATAAGGTGCGCAGCACCGCATCCGGCAACGCATATGAATATCCTCCTTAG |
| oJGB417 | TACAAGACACGTGCTGAAGTC |
| oJGB418 | TGCTAGTTGAACGCTTCCATC |
| oMKG71 | CGGTGCCCTGAATGAACTGC |
| oPM319 | CTTGTTGGATCAGACCGGAAAATG |
| oPM320 | TGGCAACGCTATCCTTTTGTCA |
| oPM407 | GCGGCAGGCAATGTGGTA |
| oPM409 | CCTTTGAGCGTGTGGTGA |
| oPM411 | CAGCGGCTACAAGCTCGG |
| oAS84 | CCCGTCTCGATCTGGTGCAG |
| oAS85 | TTGCTGCAAAAATCGCCCAAG |
| oAS132 | CCCGTCTCGATCTGGTGCAGAAGAAAGGCGGAAGTTGCCTCTGGCGAGCAGGCTGGCTCCGCTGCTGG |
| oAS133 | TTGCTGCAAAAATCGCCCAAGTCGCTATTTTTAGCGCCTTTCACAGGTATCATATGAATATCCTCCTTAG |
| oAS145 | AGGGCGTTGCGCTTCTCCGCCCAACCTATTTTTACGCGGCGGTGCCAATGAGTAAAGGAGAAGAACTT |

**Table S4 Cell doubling times of labelled strains.** Values expressed are the means of three independent replicates, with standard deviation and standard errors indicated.

| Strain (Relevant genotype) | LB medium | | | Minimal medium | | |
|---|---|---|---|---|---|---|
| | Mean doubling time (min) | SD (min) | SEM (min) | Mean doubling time (min) | SD (min) | SEM (min) |
| WT | 23.72 | 1.11 | 0.64 | 62.8 | 17.06 | 9.85 |
| dnaQ-mCherry | 26.85 | 0.91 | 0.52 | 62.22 | 18.63 | 10.76 |
| mGFP-uvrD | 24.79 | 0.37 | 0.22 | 59.45 | 1.8 | 1.04 |
| dnaQ-mCherry mGFP-uvrD | 25.27 | 1.51 | 0.87 | 71.03 | 7.59 | 4.38 |
| ΔmutS | 24.59 | 0.55 | 0.32 | 107.91 | 13.96 | 8.06 |
| ΔuvrA | 27.99 | 0.23 | 0.13 | 92.25 | 5.86 | 3.38 |
| ΔrpoB*35 | 33.98 | 0.17 | 0.1 | 83.11 | 30.83 | 17.8 |
| Δrep | 24.3 | 2.03 | 1.17 | 62.74 | 1.2 | 0.7 |
| dnaQ-mCherry mGFP-uvrD ΔmutS | 26.85 | 0.68 | 0.39 | 110.84 | 34.91 | 20.16 |
| dnaQ-mCherry mGFP-uvrD ΔuvrA | 24.13 | 0.56 | 0.33 | 91.8 | 5.18 | 2.99 |
| dnaQ-mCherry mGFP-uvrD ΔrpoB*35 | 32.81 | 0.7 | 0.4 | 99.11 | 78.68 | 45.42 |
| dnaQ-mCherry mGFP-uvrD Δrep | 26.54 | 1.9 | 1.1 | 62.17 | 2.98 | 1.72 |

**Supplementary figures**

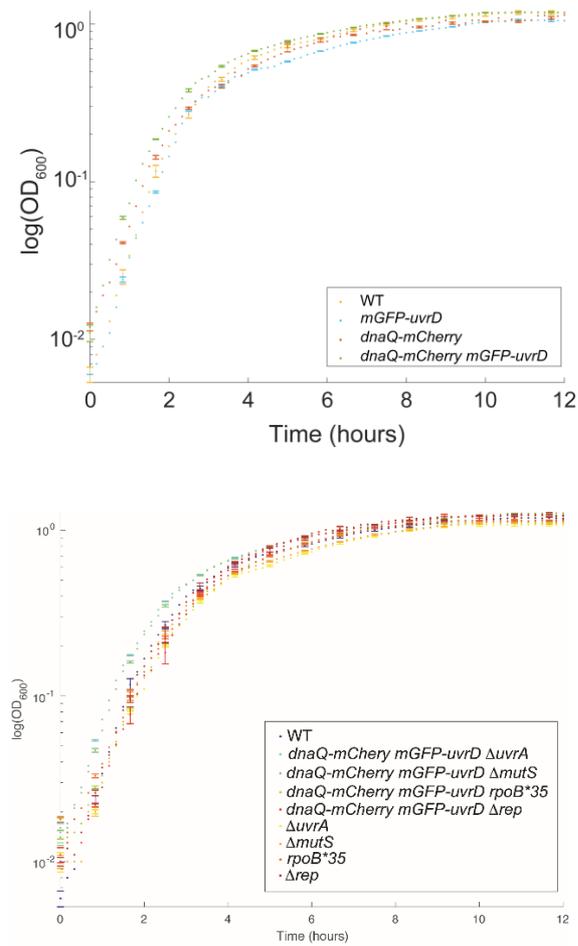

**Supplementary Figure S1. Growth curves in LB medium.** Cultures were grown in LB as described in the text. The $OD_{600}$ values are plotted on a log scale on the vertical axis with a linear scale of time in hours on the horizontal axis. The genotypes of the strains are indicated in boxes within the figures. SD error bars (shown just on every 5th consecutive point for clarity here), taken from N=3 replicate cultures. Plots in panel A are growth curves of single and dual labelled wild-type strains, while those in panel B are dual labelled and unlabelled mutant strains. The unlabelled wild-type strain is included in all panels as a reference.

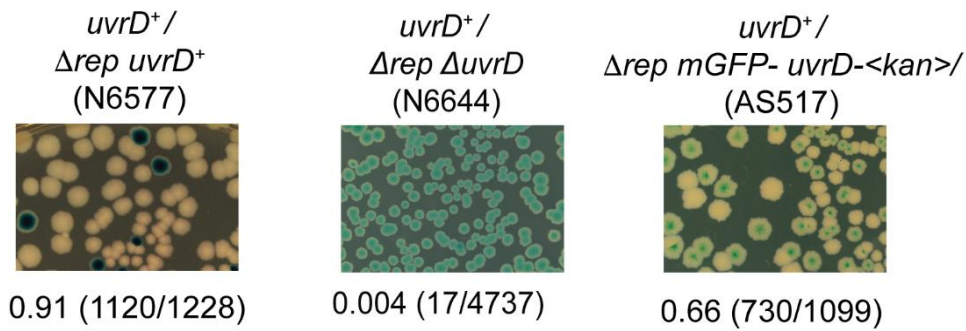

**Supplementary Figure S2: Testing of mGFP-UvrD fusions for retention of function.** We transformed the strain carrying a chromosomal *mGFP-uvrD* allele with a derivative of pRC7 carrying a wild-type *uvrD* allele (pPM407), which is an unstable low copy plasmid that also carries the *lacZYA* genes (3). Presence of this plasmid confers blue colour to the colonies on plates containing Xgal when strains are chromosomally deleted for the *lac* operon. In rapidly growing cells, cells require either *rep* or *uvrD* for viability, and the loss of both is lethal (4, 7). Thus, *Δrep* cells with a functional *uvrD* allele are viable and can lose pRC7*uvrD*, evidenced by appearance of white colonies on LB Xgal IPTG plates, but cells lacking *uvrD* function cannot survive upon loss of the plasmid resulting in recovery of only blue colonies. The *Δrep mGFP-uvrD* cells produced white colonies on Xgal media, indicating that the *mGFP-uvrD* fusion is functional.

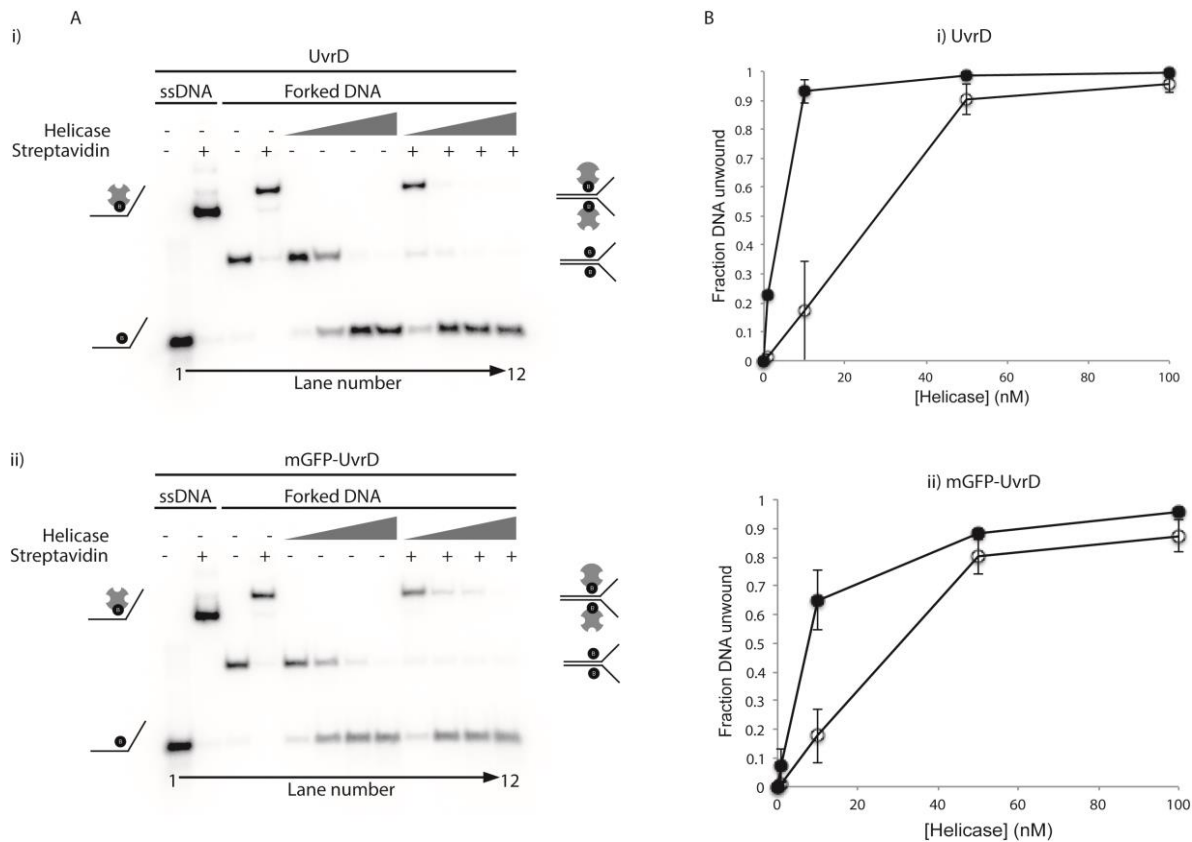

**Supplementary Figure S3: mGFP-UvrD fusion is functional *in vitro*.** (A) Representative native polyacrylamide TBE gels showing (i) UvrD and (ii) mGFP-UvrD unwinding of forked DNA containing biotin on both strands within the duplex region (8, 9). Lanes 1-4 contain markers indicating the position of single stranded or forked DNA +/- streptavidin as indicated. Lanes 5-12 contain the products of unwinding the forked DNA +/- streptavidin by the indicated helicase at 1, 10, 50 and 100 nM. (B) Quantification of the unwinding of the forked substrate in the absence of (open circles) and presence of (closed circles) streptavidin by the indicated helicases. Data points shown are averages of band intensities (taken from two gels in UvrD and four gels in mGFP-UvrD) with standard errors indicated by bars on the points.

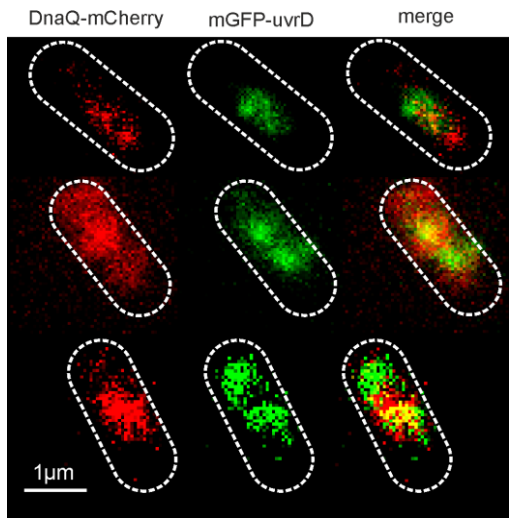

**Supplementary Figure S4:** Dual colour Slimfield images of mGFP-UvrD:DnaQ-mCherry, scale bar 1 micron. Segmented outlines of cell bodies indicated (white dash).

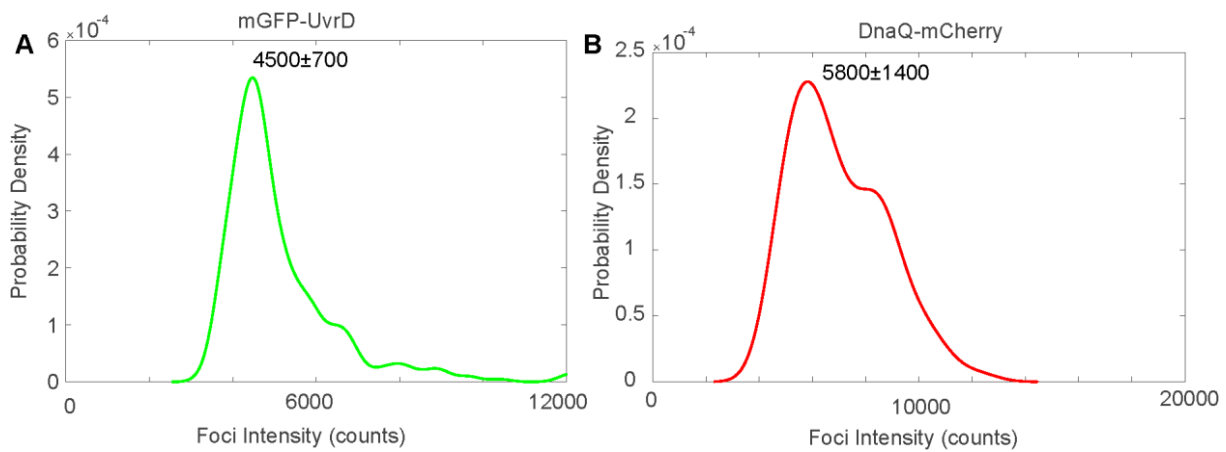

**Supplementary Figure 5: Brightness of single mGFP and mCherry molecules** A. and B. Characteristic intensity distributions rendered as kernel density estimates of single mGFP-UvrD and DnaQ-mCherry. Peak±full width at half maximum indicated. Distributions calculated from the tracked foci intensity distributions from the end of the photobleach process such that only single fluorophore molecules are detected. Number of molecules per foci before bleaching is determined by dividing the initial foci intensity by these values for the equivalent fluorophore.

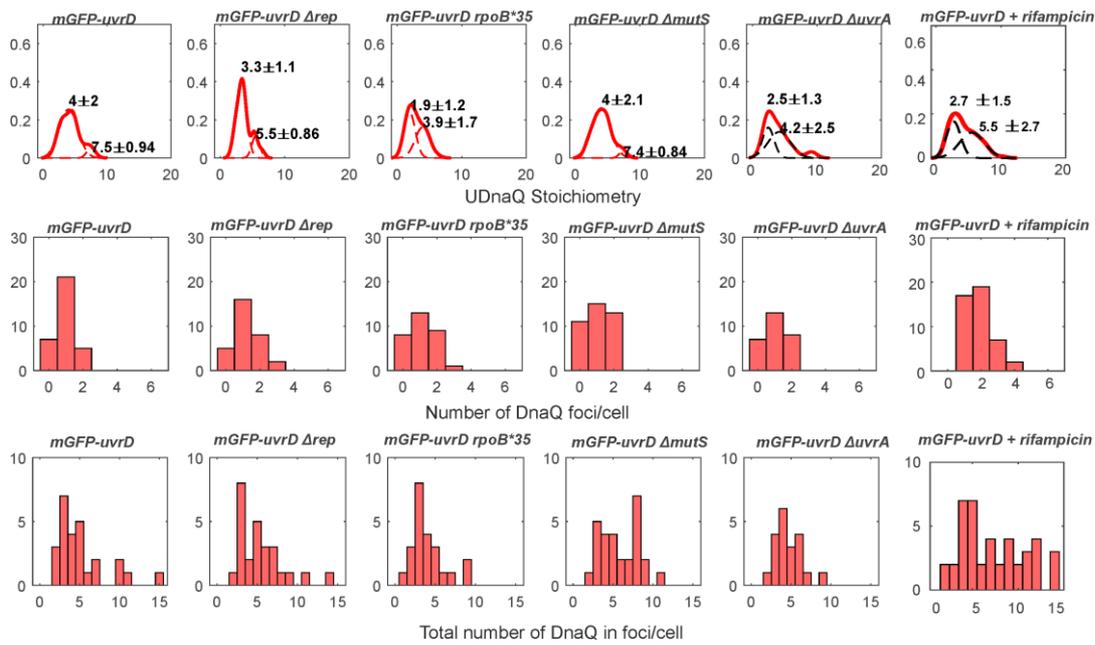

**Supplementary Figure S6: DnaQ tracking characterization.** Top row: Distribution of DnaQ foci stoichiometry as a kernel density estimate for all strains and conditions (red) with two Gaussian fit overlaid as black dotted lines. Peak ± half width at half maximum shown above. Middle row: Corresponding histogram of number of foci detected per cell. Bottom row: histogram of the total amount of DnaQ in foci per cell. N=30-50 cells per strain or condition.

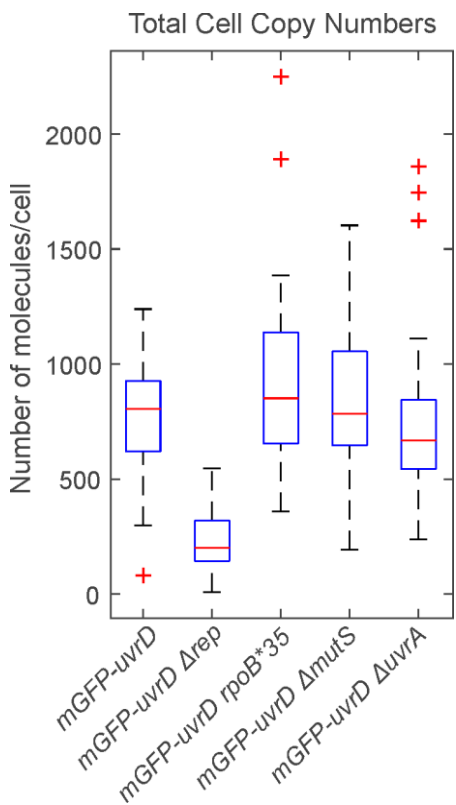

**Supplementary Figure S7:** Boxplot of the total number of mGFP-UvrD molecules per cell, estimated by numerical integration of the whole cell fluorescence. Median is shown as red

line, bottom and top of the blue box mark the 25$^{th}$ and 75$^{th}$ percentiles and whiskers extend to the most extreme points not considered outliers (2.7 standard deviations covering 99.3% of normally distributed data) with outliers beyond this shown as red +. N=30 cells per strain.

A

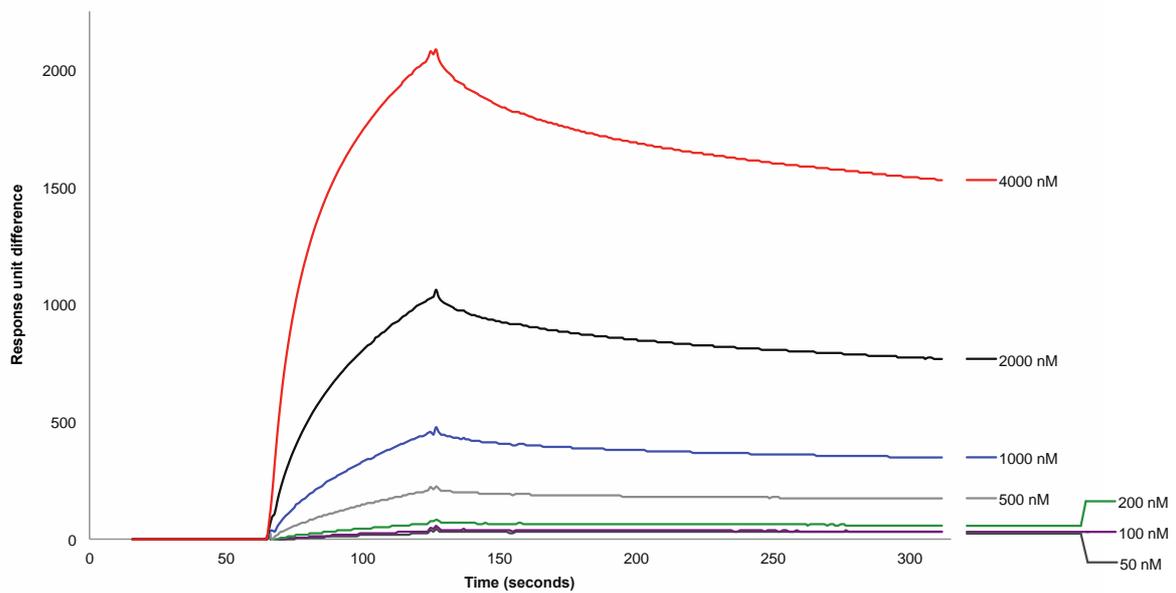

B

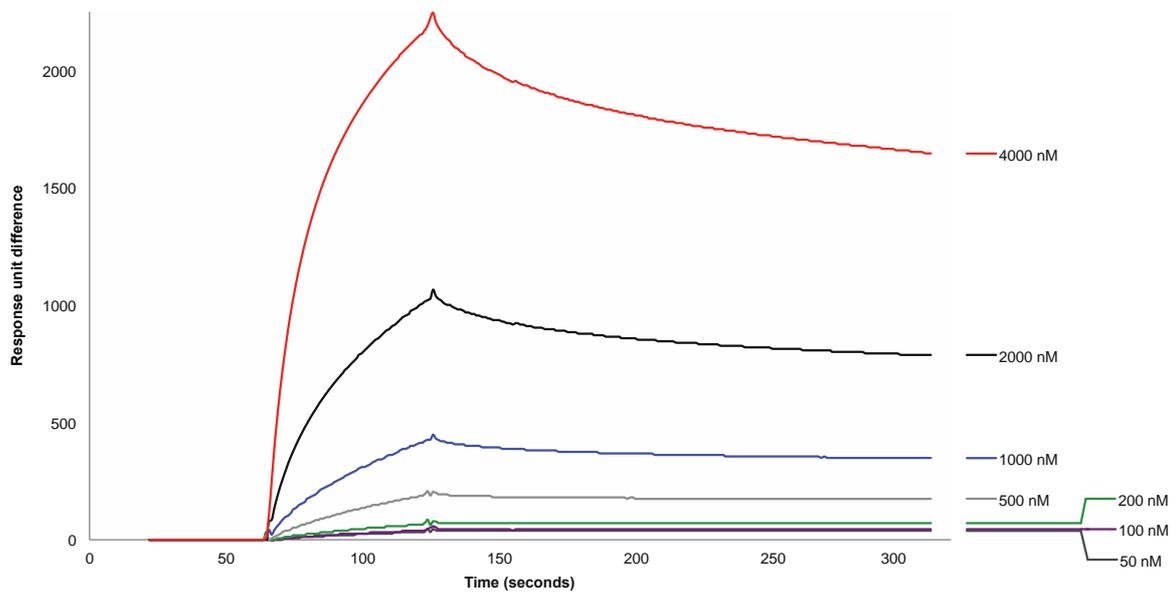

**Supplementary Figure 8: Surface plasmon resonance measurements of helicase interactions with Tau**. Binding of Tau to (A) biotinylated UvrD, 6800 RU and (B) biotinylated PcrA, 7050 RU immobilised on a streptavidin chip. Concentrations of Tau as shown.

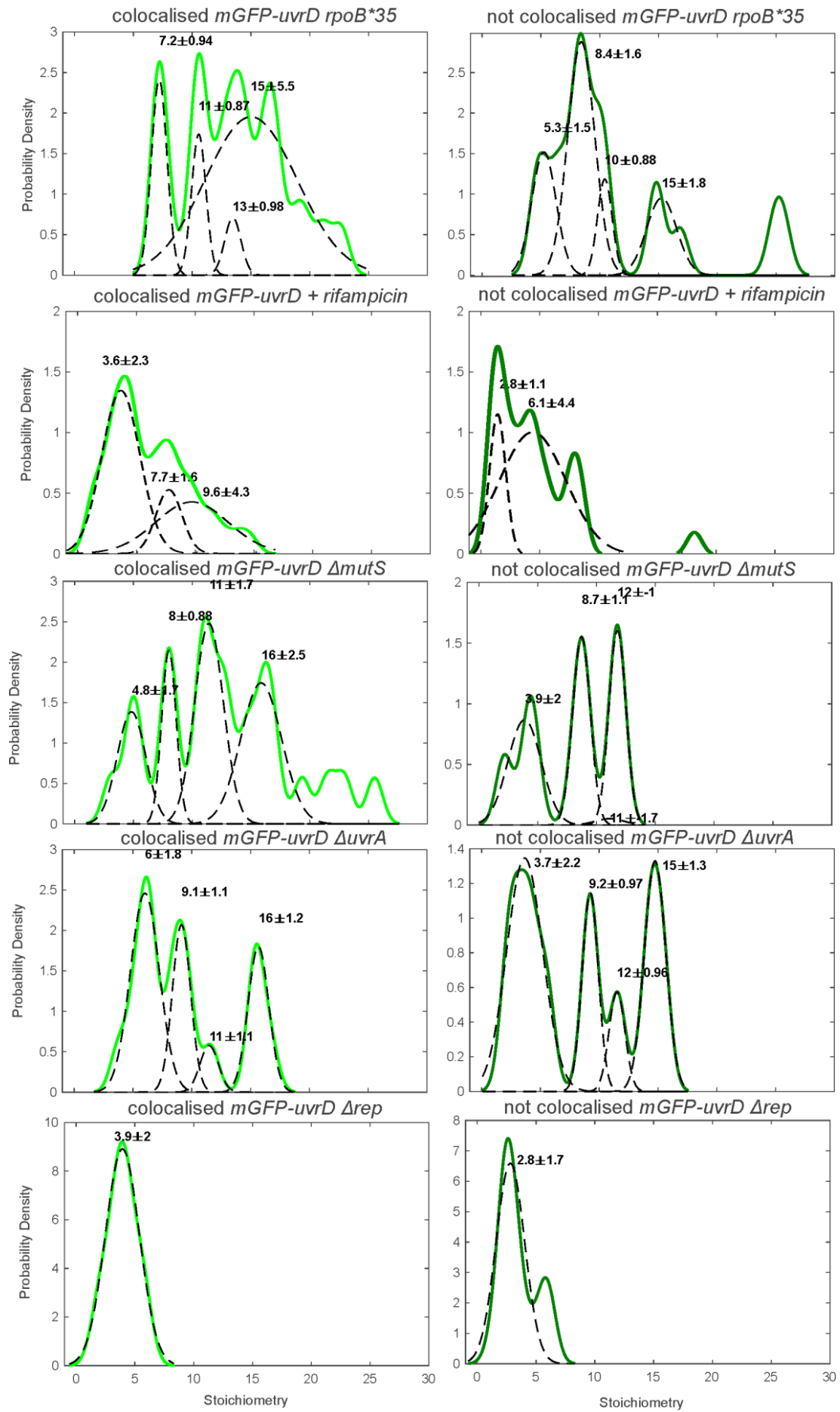

**Supplementary Figure 9:** Perturbed UvrD stoichiometry. The distribution of UvrD foci stoichiometry rendered as a kernel density estimate for foci colocalised (Left) and non-colocalised (Right) with DnaQ foci. Distributions were fitted with multiple Gaussian fits (black dotted lines) with the peak value ± half width at half maximum indicated above each peak.

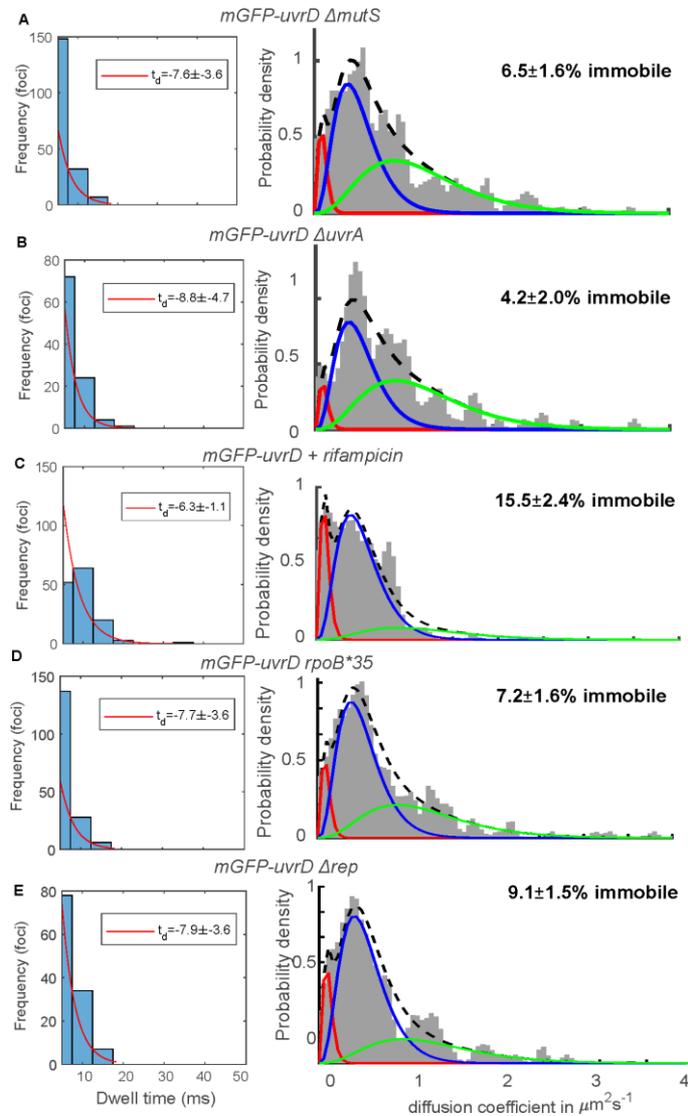

**Supplementary Figure 10:** UvrD perturbed dynamics. A-E Left, distribution of time UvrD and DnaQ foci were colocalised (blue) fit with an exponential (red) to yield a characteristic dwell time. Right, distribution of microscopic diffusion coefficients of UvrD foci (grey) fitted with a three state gamma distribution model containing a relatively immobile population with $D = 0.1$ µm$^2$/s (red), a transiently immobile population, $D = 0.5$ µm$^2$/s (blue) and a mobile population, $D=1.2$ µm$^2$/s (green). Errors represent 95% confidence intervals. N~300 foci/condition.